\documentclass{aa}
\usepackage{natbib}
\usepackage{amsmath}
\bibpunct{(}{)}{;}{a}{}{,}

\begin{document}

\title{The non--perturbative regime of cosmic structure formation}

\bigskip\medskip

\author{Thomas Buchert}

\authorrunning{Buchert}

\institute{Arnold Sommerfeld Center for Theoretical Physics, 
Ludwig--Maximilians--Universit\"at, Theresienstr. 37, D--80333 M\"unchen, Germany 
$\;-\;$\email{buchert@theorie.physik.uni-muenchen.de}
}
\date{\today}

\abstract{This paper focusses on the barely understood gap between the weakly nonlinear
regime of structure formation and the onset of the virialized regime. While the former 
is accessed through perturbative calculations and the latter through virialization
conditions  incorporating dynamical stresses that arise in collisionless self--gravitating 
systems due to velocity dispersion forces, the  addressed regime can only be understood 
through non--perturbative models. We here present an exact Lagrangian integral that 
provides a tool to access this regime. We derive a transport equation for the
peculiar--gravitational field strength and integrate it along comoving trajectories of fluid 
elements. The so--obtained integral provides an exact expression that solves the 
longitudinal gravitational field equation in general. We argue that this integral
provides a powerful approximation beyond the Lagrangian perturbative regime, 
and discuss its relation to known approximations, among them Lagrangian perturbation 
solutions including the {\it Zel'dovich approximation} and approximations for adhesive
gravitational clustering, including the {\it adhesion approximation}. 
Furthermore, we propose an iteration scheme for a systematic analytical and numerical
construction of trajectory fields. The integral may also be employed to improve
inverse reconstruction techniques.

}
\maketitle

\section{Introduction}
\label{sec:intro}

Both the analytical understanding and the numerical simulation of the formation of
cosmic structure in the Universe have substantially 
advanced in recent years. 
Early results in the development of both approaches and their comparison has improved our 
understanding of the building blocks of large--scale structure. Both focussed
on modelling {\it dark matter} in terms of a {\it Newtonian 
dust continuum}, but recent developments are heading in somewhat different directions.
Numerical attempts are directed towards improving
spatial resolution, understanding the force distribution on a lattice, 
and incorporating new physics for the modelling of structure formation on
galaxy halo scales (e.g., Bagla \& Padmanabhan 1997; Gabrielli et al. 2006;
Bertschinger 1998; Shirokov \& Bertschinger 2006)
and for the modelling of hydrodynamical effects (Steinmetz 2003).
Analytical improvements have still followed the route of understanding 
structure formation in a self--gravitating dust continuum, but also attempt to access
small--scale structure by incorporating additional forces in the Euler equation that arise
from kinetic theory (e.g., Ma \& Bertschinger 2004; Buchert \&  Dom\'\i nguez 2005).
In this work we discuss a regime that is difficult to understand and to model
using both numerical and analytical strategies. The analytical insight
will also shed light on and provide tools for numerical 
and semi--analytical techniques.

In Sect.~\ref{sec:nonpert} we recall the basic equations, 
as well as cosmological models
that have been obtained previously and that help to locate the non--perturbative regime
from the analytical point of view.
In Sect.~\ref{sec:integral} we then
formulate the key equation of the present work and integrate this equation exactly 
along the flow lines of continuum elements. The implications of this result are investigated in
Sect.~\ref{sec:implications}, and in Sect.~\ref{sec:summary} 
we give a discussion summary.

In this work we do not provide a comprehensive reference list, because the reader may find 
systematic derivations of the equations of  Sect.~\ref{sec:nonpert} as well as a 
substantial list of references in a recent paper
on the current status of analytical models (Buchert \& Dom\'\i nguez 2005). 
We first work in Eulerian coordinates that are comoving with a given 
reference {\it Hubble flow},
i.e. a homogeneous--isotropic solution of the Euler--Newton system of equations,
as is widely used in cosmology. We denote them by ${\bf q} = {\bf x}/a(t)$, where $\bf x$ are
the Eulerian coordinates, $a(t)$ a solution of Friedmann's differential equation, and
$H:=\dot a / a$ is Hubble's function. 
Later, we  move to the Lagrangian picture of fluid motion and
represent the comoving trajectory field of continuum elements by ${\bf q}={\bf F}({\bf X},t)$, where 
${\bf X}$ are the Lagrangian coordinates, which index fluid elements. 
The velocity field is split into a Hubble--velocity and a peculiar--velocity,
${\bf v}={\bf v}_H + {\bf u}$, the acceleration field into a Hubble--acceleration and 
a peculiar--acceleration ${\bf g}={\bf g}_H + {\bf w}$, the density field into a background density
and a density contrast $\varrho = \varrho_H (1 + \delta)$.  
The total (Lagrangian) time--derivative is denoted by 
$\frac{d}{dt} = \partial_t \vert_{\bf x} + {\bf v}\cdot\nabla_{\bf x}= 
\partial_t \vert_{\bf q}+ {\bf u}/a \cdot \nabla_{\bf q}$, and sometimes by an
overdot. The nabla--operator with respect to Lagrangian coordinates is denoted by 
$\nabla_{0}$ and sometimes by $\nabla_{\bf X}$.

\section{The non--perturbative regime}
\label{sec:nonpert}

In this section we recall the basic equations in a form that is helpful for our 
purpose and then review the known cosmological model equations within the presented
framework and discuss their limits with regard to non--perturbative effects in the non--linear
regime.

\subsection{Equations for the non--perturbative regime}
\label{subsec:nonpert_equations}

For the basic system of equations, derived from coarse--graining the Newtonian kinetic equations
for a system of N self--gravitating particles, we refer the reader to our recent paper
(Buchert \& Dom\'\i nguez 2005). We work below with hydrodynamical equations 
that can also be
found in many cosmology textbooks and review papers (e.g., Peebles 1980, 
Binney \& Tremaine 1987, Sahni \& Coles 1995,
Ehlers \& Buchert 1997, Bernardeau et al. 2002).

We start with recalling equations that describe the evolution of the 
density $\varrho$, the peculiar--velocity $\bf u$, and the peculiar--acceleration or the
peculiar--gravitational field strength $\bf w$,  obtained by coarse--graining the kinetic equations
of N self--gravitating particles in real space and velocity space and then forming velocity 
moments to obtain a hydrodynamical description in Eulerian space.
The zeroth velocity moment is the {\it continuity equation},
\begin{subequations}
\begin{equation}
\label{continuity}
\frac{d}{dt}\varrho + 3H \varrho + \frac{1}{a}\nabla_{\bf q}\cdot {\bf u}\;=\;0\;\;.
\end{equation}
The first velocity moment provides the evolution equation for the peculiar--velocity:
\begin{equation}
\label{eulerjeans}
\frac{d}{dt}{\bf u} + H {\bf u} \;=\; {\bf w} + \frac{1}{\varrho} \left({\bf F}-\frac{1}{a}
 \nabla_{\bf q} \cdot \boldsymbol{\Pi}\right) \;\;.
\end{equation}
In the above equation, the peculiar--gravitational field strength ${\bf w}$ is constrained by
the {\it Newtonian field equations}:
\begin{equation}
\label{fieldequations}
\nabla_{\bf q} \times {\bf w} \;=\;{\bf 0}\;\;\;;\;\;\;\nabla_{\bf q}\cdot{\bf w}
\;=\;-4\pi G \varrho_H a \delta\;\;.
\end{equation}
\end{subequations}
The force $\bf F$ represents deviations from mean field gravity (modelled by 
Eqs.~(\ref{fieldequations})), and the symmetric tensor field
$\boldsymbol{\Pi}$ represents forces due to velocity dispersion. For ${\bf F}={\bf 0}$, 
Eq.~(\ref{eulerjeans}) is known as the {\it Euler--Jeans equation} in stellar system theory. 
The above set of equations truncates the velocity moment hierarchy, so we need models
for the fields $\bf F$ and $\boldsymbol{\Pi}$ 
(e.g., as functionals of the other fields) to close the system
of equations. The simplest truncation is to neglect deviations from mean field gravity and
velocity dispersion altogether and study the evolution of a {\it dust continuum}. More general
models for {\it adhesive gravitational clustering} including those forces are investigated
in (Buchert \& Dom\'\i nguez 2005).

In this work we focus on the (in the above framework) 
general evolution equation for the peculiar--gravitational
field strength (Buchert 1989)
\begin{eqnarray}
\label{w_evolution}
\frac{d}{dt}{\bf w} + 2H{\bf w} - 4\pi G \varrho_H {\bf u} \;=\; 
\boldsymbol{\cal R}\;\;; \nonumber\\
\qquad\boldsymbol{\cal R}: = \frac{1}{a} \left[\,({\bf u} \cdot \nabla_{\bf q}) 
{\bf w} -{\bf u}(\nabla_{\bf q} \cdot {\bf w}) + 
\nabla_{\bf q} \times {\bf T}\,\right] \;\;,
\end{eqnarray}
which can be obtained from Eq.~(\ref{continuity}) by inserting the field equations
(\ref{fieldequations}) and formally integrating the divergence. The resulting vector field of
integration $\bf T$ can be determined as follows.
Acting with $\nabla_{\bf q}\times$ on
Eq.~(\ref{w_evolution}) and subjecting $\bf T$ to the {\it Coulomb gauge condition},
$\nabla_{\bf q} \cdot {\bf T} = 0$, one finds:
\begin{eqnarray}
\label{Tconstraint}
{\Delta}_{\bf q} {\bf T} & = & 4\pi Ga \nabla_{\bf q} \times (\varrho {\bf u})
= 4\pi G a \left[ \varrho \nabla_{\bf q} \times {\bf u} + 
\nabla_{\bf q}\varrho \times {\bf u} \right] 
\nonumber \\
& = & (4\pi Ga \varrho_H - \nabla_{\bf q} \cdot {\bf w}) \, 
\nabla_{\bf q} \times {\bf u}
 + {\bf u}\times {\Delta}_{\bf q} {\bf w}\; .
\end{eqnarray}
That is, $\bf T$ is the vector potential (up to an unimportant factor)
of the {\it peculiar--current density} ${\bf j}: =\varrho {\bf u}$:
\begin{equation}
\label{currentdensity}
4 \pi G a {\bf j}\;=\;-\nabla_{\bf q} \times {\bf T} - 
\nabla_{\bf q} \,\frac{1}{a}\partial_t \vert_{\bf q}\left[\,a\phi\,\right]\;\;,
\end{equation} 
with $\phi$ denoting the scalar peculiar--gravitational potential.
(Note that Eq.~(\ref{currentdensity}) is equivalent to Eq.~(\ref{w_evolution}) by 
employing the definitions $\frac{d}{dt}{\bf w}= \partial_t \vert_{\bf q} {\bf w}+ 
{\bf u}/a \cdot\nabla_{\bf q}{\bf w}$ and ${\bf w}=:-\frac{1}{a}
\nabla_{\bf q} \phi$.)
If we require $\bf T$ to vanish, then Equation~(\ref{Tconstraint}) shows that, 
for irrotational flows, the mean flow follows the gradient of the density 
field\footnote[3]{If such conditions
  are imposed on the problem, $\bf T$ becomes a harmonic vector function
  that can be set to zero for periodic boundary conditions, since 
  harmonic functions are then spatially constant and can be set to zero due
  to the invariance of the basic equations with respect to spatially
  constant translations.}). 

We thus have obtained a new set of equations that is equivalent to
the set (\ref{continuity},\ref{fieldequations}), and we
can write it in a form that reminds us of Maxwell's equations:
\begin{eqnarray}
&\displaystyle\frac{1}{a} \nabla_{\bf q} \times {\bf w} = {\bf 0}\qquad\qquad\quad\;
&\;\;\; \displaystyle\frac{1}{a} \nabla_{\bf q} \cdot {\bf w} = - 4 \pi G \varrho_H \delta\;\;;
\nonumber\\
& \displaystyle\frac{1}{a} \nabla_{\bf q} \times {\bf T} = \partial_t \vert_{\bf q}
{\bf w}- 4\pi G {\bf j}
&\;\;\; \displaystyle\frac{1}{a} \nabla_{\bf q} \cdot {\bf T} =  0\;\;.
\end{eqnarray}

\vspace{5pt}

For the purpose of constructing models for cosmic structure formation, one usually employs the 
equations (\ref{continuity},\ref{eulerjeans},\ref{fieldequations}) and approximates the fields
$\bf F$ and $\boldsymbol{\Pi}$. A commonly used approximation is to set $\bf F$ equal to 
zero and ``masking''
$\boldsymbol{\Pi}$ by its isotropic contribution, $\Pi_{ij}=:p \delta_{ij}$, together with
a dynamical equation of state $p=\beta (\varrho)$ (Buchert \& Dom\'\i nguez 1998).  
For such a closure approximation, we obtain the starting equations for a large
set of approximations that have been discussed in the cosmological literature. 

To recall
these approximations, let us now move to the Lagrangian picture of fluid mechanics, where
the Eulerian positions are viewed as a field variable depending on the initial position
vectors $\bf X$ (the Lagrangian coordinates) and the time $t$, ${\bf q}={\bf F}({\bf X},t)$.
The trajectory field $\bf F$ represents the integral curves
of the scaled peculiar--velocity field, ${\bf u} = a {\dot{\bf F}}$. We derive equations
for the comoving displacement field ${\bf P}:={\bf F}-{\bf X}$. First, we
formally integrate Eq.~(\ref{w_evolution}) with respect to the time and 
obtain (Buchert \& Dom\'\i nguez 2005):
\begin{subequations}
\begin{equation}
\label{w_general}
{\bf w}\;=\; \frac{\bf W}{a^2} + 4 \pi G \varrho_H a {\bf P}({\bf X},t) +
\frac{1}{a^2} \int_{t_0}^t  dt' \; a^2(t') \boldsymbol{\cal R}({\bf X},t')\;\;,
\end{equation}
\begin{equation}
{\rm with}\qquad
{\bf w}({\bf X},t_0) = : {\bf W}({\bf X})\;\;\;,\;\;\;{\bf P}({\bf X},t_0):={\bf 0}\;\;\;.
\qquad\qquad
\end{equation}
\end{subequations}
The integration constant has been chosen such that the Lagrangian
coordinates coincide with the comoving Eulerian coordinates at initial time
\footnote[4]{
Alternatively, we could define initial displacements through  
$4 \pi G\varrho_H (t_0) {\bf P}({\bf X}, t_0) = {\bf w}({\bf X}, t_0)$.
This means either that we have assumed
quasi--homogeneity, $\varrho ({\bf X},t_0) \approx \varrho_H (t_0)$ or, else, 
we have adopted a particular choice of Lagrangian coordinates. 
With this assumption we get rid of physically
unimportant constant terms, which is sometimes useful, but will not be used here
to avoid confusion with the exact property of the expressions derived in this paper.
(For details on the
possibility and advantage of this choice, 
see Adler \& Buchert 1999, App. A.)}. 
Notice that Eq.~(\ref{w_general}) is general and does not depend on the 
specific models for the forces in Eq.~(\ref{eulerjeans}).
Inserting this general expression for $\bf w$ into Eq.~(\ref{eulerjeans}) and using 
${\bf u} = a{\dot{\bf P}}$  yields an evolution equation for the displacement field $\bf P$,
which we simplify -- for the sake of a more transparent discussion -- 
to the case ${\bf F}={\bf 0}$ and 
an isotropic velocity dispersion tensor (representing the dynamical pressure by an equation
of state of the form $p= \beta(\varrho)$); this restriction implies:
\begin{equation}
\label{isotropy}
-  \frac{1}{a^2 \varrho} \nabla_{\bf q} \cdot \boldsymbol{\Pi} \; =\;
 \frac{L_J^2 (\varrho)}{a^3} {\Delta}_{\bf q} {\bf w} \;\;.
\end{equation}
Above, we have introduced a density dependent {\it Jeans' length} as the product of the {\it speed
of sound} $c_s$ and the {\it local free--fall time} $t_F$: 
\begin{equation}
\label{jeanslength}
L_J : = \sqrt{\frac{\beta' (\varrho)}{4\pi G\varrho}}=
c_s t_F\;,\;c_s^2 : = \frac{dp}{d\varrho}=\beta'(\varrho)
\;,\;t_F := \frac{1}{\sqrt{4\pi G\varrho}}.
\end{equation}
For this case we obtain the following evolution equation:
\begin{eqnarray}
\label{P_general}
\ddot{\bf P} + 2 H \dot{\bf P} - 4 \pi G \varrho_H {\bf P} \;=\;\nonumber\\
\qquad \frac{L_J^2 (\varrho)}{a^3} {\Delta}_{\bf q} {\bf w}
+ \frac{1}{a^2} \int_{t_0}^t \;\; dt' \; a^2(t') \boldsymbol{\cal R}({\bf X},t')\;\;.
\end{eqnarray}

\subsection{A list of known cosmological model equations}
\label{subsec:nonpert_models}

From Eq.~(\ref{P_general}) we are able to infer a list of approximation 
assumptions that have been used for the construction of cosmological evolution models:
\begin{itemize}
\item[1]
Setting $L_J =0$ formally we obtain the {\it dust model} without any restriction. 

\item[2]
Neglecting the residual vector field $\boldsymbol{\cal R} = {\bf 0}$, 
we obtain for  $L_J =0$ the equation for the longitudinal part of Lagrangian first--order 
perturbations (the transversal part is hidden in $\boldsymbol{\cal R} = {\bf 0}$). 
The peculiar--gravitational field strength is then, in view of Eq.~(\ref{w_general}), 
approximated through the relation: 
\begin{equation}
\label{w_approx}
{\bf w} = \frac{\bf W}{a^2}+4\pi G \varrho_H a{\bf P}\;\;.
\end{equation}

\item[3]
Also neglecting the residual vector field $\boldsymbol{\cal R} = {\bf 0}$,
we obtain for $L_J \ne 0$ together with the valid relation (\ref{w_approx}):
\begin{equation}
 \frac{L_J^2 (\varrho)}{a^3} {\Delta}_{\bf q} {\bf w} \;=\;
 \frac{\varrho_H \, \beta'(\varrho)}{a^2 \varrho} {\Delta}_{\bf q} {\bf P} \;\;.
\end{equation}
The coefficient in front of the (Eulerian) Laplacian of $\bf P$ is density--independent for
the special choice $p=\kappa \varrho^2$, $\kappa =
const.$, which corresponds to the assumption of an isolated and  `virialized' fluid element, 
{\it cf.} Buchert \& Dom\'\i nguez 2005, Appendix C).
For this example we obtain an evolution equation of the form:
\begin{equation}
  \label{P_approxEJN}
{\ddot{\bf P}} + 2 H {\dot{\bf P}} - 4 \pi G \varrho_H {\bf P} 
  = \frac{2 \kappa \varrho_H}{a^2} {\Delta}_{\bf q} {\bf P}\;\;.
\end{equation}
Note that the above model equation is nonlinear in Eulerian space (due to the convective
non--linearities hidden in the overdot) and in Lagrangian space (the Eulerian Laplacian,
if transformed to Lagrangian coordinates is non--linear).
It is of a {\it hybrid} Lagrangian/Eulerian type and has been suggested in
Buchert \& Dom\'\i nguez (2005).

\item[4]
Linearization of the convective non--linearities in Eq.~(\ref{P_approxEJN})
yields the Eulerian linear approximation (e.g., Peebles 1980).
For this, note that linearization of the general integral for the density (see 
Eq.~(\ref{densityintegral}) below) gives $\delta = - \nabla_{0} \cdot {\bf P}$. 

\item[5]
By linearizing the Eulerian Laplacian in Eq.~(\ref{P_approxEJN}), now in the Lagrangian
frame, i.e., 
retaining only the zero--order Lagrangian term ${\bf q} = {\bf X}+{\bf P}\;\approx\;{\bf X}$,
we recover the Lagrangian linear approximation for a medium supported by a 
dynamical pressure (Adler \& Buchert 1999). 

\item[6] Restricting the latter model further by assuming that ${\bf u} \propto {\bf w}$, 
with the function of proportionality taken from the Eulerian linear approximation for 
a dust continuum, we obtain
the  standard {\it adhesion approximation} (Gurbatov et al. 1989).

\item[7]
The {\it adhesion approximation} reduces, for $L_J = 0$, to 
the {\it Zel'dovich approximation} (Zel'dovich 1970, 1973).

\end{itemize}

\subsection{Discussion of models}
\label{subsec:nonpert_discussion}

While models 2 and 4--7 are perturbative, model 3
already describes the non--perturbative regime concerning the dynamical stress tensor.
This non--perturbative approximation
extrapolates the Lagrangian linear model by 
the replacement $\Delta_{\bf X} \rightarrow 
\Delta_{{\bf X} + {\bf P}}$.

There are other models that  are not covered by this list, notably higher--order
perturbative approximations. For example, in the Lagrangian perturbation framework
that contains the {\it Zel'dovich approximation} as a special first--order solution 
in a subclass (Buchert 1989, 1992),
solutions have been derived for {\it dust} up to the fourth order. Special classes of
second--order solutions and their investigation may be found in Bouchet et al. (1992, 1995) and
Buchert \& Ehlers (1993). To the third order Moutarde et al. (1991) gave
a special solution, Buchert (1994) and Catelan (1995) investigated a large class.
Vanselow (1996) derived second-- and third--order solutions for some more general
cases, as well as a class of fourth--order solutions.  
By including isotropic stresses, Lagrangian perturbation solutions were given to first order 
(Adler \& Buchert 1999), to second order (Morita \& Tatekawa 2001; Tatekawa 
et al. 2002), and to third order (Tatekawa 2005a,b). 
All of these higher--order models can be regarded as approximating the residual vector field
$\boldsymbol{\cal R}$. However, in all of these models, the peculiar--gravitational
field strength remains smooth while crossing high--density regions, which (as
becomes clear below) points to a shortcoming of the perturbative calculations.
The difficulty in deriving a closed differential equation for the 
displacement field $\bf P$ in the highly non--linear regime lies precisely  in 
the term $\boldsymbol{\cal R}$, which is a non--local (in space and time) 
and a non--linear functional of $\bf P$; approximating $\boldsymbol{\cal R}$ perturbatively
hides an important  effect, as we shall see in detail below.

Finding an extension into the non--perturbative (both Eulerian and
Lagrangian) regimes is an involved mathematical task. Notwithstanding,
it should be attempted. For example, even the Lagrangian
perturbation approach falls short capturing the action of
multi--stream forces. This was demonstrated by comparing the statistical properties
of second-- and third--order Lagrangian perturbation solutions
with results of numerical simulations (Tatekawa 2004a,b). 
This shortcoming calls for a non--perturbative generalization.

A proper understanding of the extrapolation of the linear relationship (\ref{w_approx})
between $\bf w$ and $\bf P$ into the non--perturbative regime requires that the
residual term $\boldsymbol{\cal R}$ in Eq.~(\ref{w_general}) be analyzed in general. 
To this end we exploit the fact that
Eq.~(\ref{w_general}) holds independently of whether we are talking about {\it dust} or
a general dispersion--supported system. It is possible to find exact integrals of the 
general equation (\ref{w_general}),
which will also help us to understand the quality of the relationship (\ref{w_approx}) --
lying at the basis of most currently known models -- 
without involving higher--order perturbation analysis.

Finally, although virialized states can be 
understood through the tensor virial theorem, eventually including surface terms to account
for the non--isolated state of gravitational systems, there are signatures possibly imprinted
onto the phase space distribution during the non--perturbative regime. This 
distribution may have a ``relaxed'' global shape, but its internal structure 
will probably appear structured, rather than completely smooth, e.g.  as a result of a hierarchy of
embedded (smoothed) caustics (Ed Bertschinger, {\it priv. comm.}).

\section{Exact integral for the peculiar--gravitational field strength}
\label{sec:integral}

In this section we look at the general equations and combine them in order to obtain a transport
equation for the peculiar--gravitational field strength. We then integrate the transport 
equation exactly with the help of a (sufficient) restricting condition. 
With this result we are able to calculate the gravitational field strength 
for a given family of trajectories,  and to solve the longitudinal field 
equation $\nabla_{\bf q} \cdot {\bf w} = -4\pi G \varrho_H a \delta$ in general.
We later demonstrate that this integral can be exploited to obtain 
a powerful approximation for the non--perturbative regime of structure formation.

\subsection{Transport equation for the peculiar--gravitational field strength}
\label{subsec:integral_transport}

We start with Eq.~(\ref{w_evolution}) and employ the vector identity
\begin{eqnarray}
\label{identity}
({\bf u} \cdot \nabla_{\bf q}){\bf w} - {\bf u} (\nabla_{\bf q} 
\cdot {\bf w})\;=\nonumber\\
\qquad\left[\, ({\bf w} \cdot \nabla_{\bf q}){\bf u}  -
{\bf w} (\nabla_{\bf q} \cdot {\bf u})\, \right]+\nabla_{\bf q}  \times ({\bf w} \times {\bf u})
\end{eqnarray}
to find the modified evolution equation:
\begin{eqnarray}
\label{w_evolutionmodified}
\frac{d}{dt}{\bf w} + 2H{\bf w} - 4\pi G \varrho_H {\bf u} \;=\;\qquad\nonumber\\
\qquad\frac{1}{a} \left[\,\left(\,{\bf w} \cdot \nabla_{\bf q}\,\right) {\bf u} -
{\bf w}\left(\,\nabla_{\bf q} \cdot {\bf u}\,\right)+
\nabla_{\bf q} \times {\bf {\tilde T}}\,\right] \;\;,
\end{eqnarray}
with a new vector potential ${\bf\tilde T}:= {\bf T} + {\bf w}\times{\bf u}$.
Computing $\;\varrho \frac{d}{dt} \left({\bf w}/{\varrho}\right)\;$, and
using the continuity equation (\ref{continuity}), we arrive at the following equation that we
may call {\it transport equation} for $\bf w$:
\begin{eqnarray}
\label{w_transport}
\frac{d}{dt}\left(\frac{\bf w}{\varrho}\right) - H \left(\frac{\bf w}{\varrho}\right)\;=\;
\qquad\nonumber\\
\qquad\left(\frac{\bf w}{\varrho}\cdot \nabla_{\bf q}\right)\,\frac{\bf u}{a} +
4\pi G \varrho_H \frac{\bf u}{\varrho} +\frac{1}{a\varrho}\nabla_{\bf q}
\times {\bf {\tilde T}}\;\;.
\end{eqnarray}
This is the key--equation of the present work.

\subsection{Integrating the transport equation}
\label{subsec:integral_integral}

For the case $\nabla_{\bf q}\times{\bf {\tilde T}}={\bf 0}$, we can
find an exact integral to the above transport equation along comoving trajectories
${\bf q}={\bf F}({\bf X},t)$ as follows.

Motivated by a recent investigation of an exact Lagrangian integral for the gravitational
field strength $\bf g$ in Newtonian gravity (Buchert 2006a), we make the following
ansatz (which generalizes the ansatz for the case of a non--vanishing background source
in the above--mentioned work):
\begin{equation}
\label{ansatzforintegral}
\frac{{\bf w} - \zeta\,  {\bf F}}{\varrho} = a \left(\,{\bf k} \cdot \nabla_{0}\,
\right)\,{\bf F}\;\;\;;\;\;\;\zeta = \zeta (t)\;\;\;;\;\;\;{\bf k} = {\bf k}({\bf X})\;\;,
\end{equation}
with the nabla--operator with respect to Lagrangian coordinates 
$\nabla_{0}$.
Performing the total time--derivative of this ansatz along the integral curves
${\bf F} = {\bf X}+{\bf P}$ of the scaled peculiar--velocity field, i.e.,
$\frac{d}{dt}{\bf F} = \frac{1}{a}{\bf u} ({\bf F},t)$, using a further identity (e.g.,
Serrin 1959),
\begin{equation}
\left(\,{\bf k} \cdot \nabla_{0}\,\right)\, \frac{d}{dt} {\bf F} \;=\; 
\left[\, ({\bf k} \cdot \nabla_{0})\,{\bf F}\, \right] \cdot 
\nabla_{\bf q} \frac{\bf u}{a}\;\;,
\end{equation}
and once more applying the identity (\ref{identity}) to the fields $\bf u$ and $\bf F$,
\begin{eqnarray}
\nabla_{\bf q} \cdot \,\lbrack \,{\bf F} (\nabla_{\bf q} \cdot {\bf u}) -
({\bf F} \cdot \nabla_{\bf q}) {\bf u} \,\rbrack \;=\;\qquad\qquad\nonumber\\
\nabla_{\bf q} \cdot \,\lbrack \,{\bf u} (\nabla_{\bf q} \cdot {\bf F}) - 
({\bf u}\cdot \nabla_{\bf q}) {\bf F} \,\rbrack \,=\,
\nabla_{\bf q} \cdot (d-1)\,{\bf u}\;,
\end{eqnarray}
(with $d$ denoting the dimension of space), we obtain an equation that we compare with the
transport equation.
We are left with the following 
conditions on the unknown function $\zeta$:
\begin{equation}
\frac{\bf F}{\varrho} \left(\,{\dot \zeta} + 2H \zeta\,\right) = {\bf 0}
\;\;;\;\;\frac{\bf u}{\varrho}\left(\,\zeta - \frac{4\pi G \varrho_H a}{d}\,\right)
= {\bf 0}\;\;.\nonumber
\end{equation}
For non--vanishing $\bf F$ and $\bf u$, the above two equations
for $\zeta$ are equivalent by virtue of ${\dot\varrho}_H = -3H \varrho_H$,
and we have determined the unkown function:  
\begin{equation}
\zeta = \frac{4\pi G\varrho_H a}{d}\;\;.
\end{equation}

Now, we can write down an exact integral ${\bf w}^{I}$
for the peculiar--gravitational field--strength. We also replace the density by its exact 
Lagrangian integral, 
\begin{equation}
\label{densityintegral}
\varrho ({\bf X},t) = \frac{\varrho({\bf X},t_0)}{J} \;=\;
\varrho_H \frac{1+\delta ({\bf X},t_0)}{J_F}\;\;;\;\;
\varrho_H =\frac{\varrho_H (t_0)}{a^3}\;\;,
\end{equation}
where
$J :=\det(f_{i|k})$ denotes the Jacobian of the transformation from ${\bf x}={\bf f}({\bf X},t)$ 
to $\bf X$, and $J_F :=\det(F_{i|k})=Ja^{-3}$ denotes the Jacobian of the transformation from 
${\bf q}={\bf F}({\bf X},t)$ to $\bf X$
with the comoving Lagrangian deformation gradient $(F_{i|k})$\footnote[5]{We denote
a spatial derivative with respect to Lagrangian coordinates by a vertical slash $|$
that commutes with the Lagrangian time--derivative.}).
The result reads
\begin{equation}
\label{w_exactintegral}
{\bf w}^{I} \; = \; 
\frac{\left(\,{\bf K} \cdot \nabla_{0}\,\right) {\bf F}}{a^2 J_F} \;+ \;
\frac{4 \pi G\varrho_H a}{d}{\bf F} \;\;;
\end{equation}
${\bf K} = \frac{\bf k}{\varrho (t_0)} = {\bf W} - \frac{4 \pi G
\varrho_{H}(t_0)}{d}{\bf X}$ is the integration constant, which was determined by
evaluating the  integral at $t=t_0$.
The integral (\ref{w_exactintegral}) explicitly depends on the dimension $d$ of the
continuum. 

\vspace{3pt}

The above integral is {\it quasi--local}, i.e. it locally represents the peculiar--gravitational
field strength through a functional of the comoving trajectory field $\bf F$, while initial
data are constructed non--locally according to the structure of the theory.

A strong property of the  integral  (\ref{w_exactintegral}) is that it solves 
the longitudinal field equation $\nabla_{\bf q}\cdot {\bf w} =-4\pi G \varrho_H a \delta$
{\it in general}.
(We can also find such an integral for the remaining field equations $\nabla_{\bf q}
\times {\bf w} = {\bf 0}$, see Sect. 5; a general integral is expected to include
non--local terms.) 

\subsection{Lagrangian framework, proof of the exact integral and its  
transformation properties}
\label{subsec:integral_proof}

Let us recall the basic equations of a Lagrangian description of fluid motion.
As shown in Buchert (2006a), the presented integral 
(\ref{w_exactintegral}) can also be obtained by simply transforming the result obtained in 
Newtonian gravity. This is expected for exact equations and expressions.
The explicit derivation in this paper shows that the approximation of a vanishing curl 
of the vector potential $\bf\tilde T$ apparently does 
not impair this property; in general, approximations to the system of equations in Newtonian
gravity do not carry over to corresponding approximations in Newtonian cosmology by a
simple transformation.

\smallskip

In Newtonian gravity
the Lagrangian description is based on integral curves
${\bf x} = {\bf f}({\bf X},t)$ of the full velocity field
${\bf v}({\bf x},t)$:
\begin{equation}
\frac{d {\bf f}}{dt} = {\bf v} ({\bf f}, t) \;\; ; \;\;
{\bf f}({\bf X}, t_0) = :{\bf X} \;\; .
\end{equation}
By introducing this family of trajectories, we can express
all Eulerian fields (e.g., the velocity ${\bf v}$, the acceleration $\bf g$, 
the density $\varrho$, and the vorticity 
$\boldsymbol{\omega}:=\frac{1}{2}\nabla_{\bf x}\times {\bf v}$)
in terms of the field of trajectories
${\bf x}={\bf f} ({\bf X},t)$ as follows:
\begin{equation}
\label{lagevolution}
{\bf v} = \dot {{\bf f}} ({\bf X},t)\;\;;\;\;{\bf g} = \ddot {\bf f} ({\bf X},t)\;\;;
\end{equation}
\begin{equation}
\label{lagintegrals}
\varrho  = \frac{\varrho_0}{J}\;\;;\;\;
\boldsymbol{\omega} = \frac{\boldsymbol{\omega}_0
\cdot \nabla_{0} {\bf f}}{J}\;\;,
\end{equation}
with the Jacobian of the transformation from Eulerian to Lagrangian coordinates
$J:= \det(f_{i | k}({\bf X},t))\;> 0$, $\varrho_0 :=\varrho ({\bf X},t_0)$,
$\boldsymbol{\omega}_0 :=\boldsymbol{\omega} ({{\bf X}},t_0)$.

Equation~(\ref{lagintegrals}) lists the known {\it Lagrangian integrals} for the density $\varrho$ and 
the vorticity $\boldsymbol{\omega}$ 
of the Euler--Newton system; i.e., they represent a Eulerian field as a functional of $\bf f$.
To transform those fields back to Eulerian space, we need the transformation
${\bf f}$ to be invertible; i.e. $J>0$, defining {\it regular solutions} 
(for more details the reader may consult the review by Ehlers \& Buchert 1997).

The Eulerian {\it field equations} are transformed into a system
of Lagrangian equations by virtue of the following transformation of the 
field strength gradient:
\begin{equation}
\label{g-gradient}
 \frac{\partial}{\partial x_j}g_i = g_{i | k} \frac{\partial}{\partial x_j}h_k = 
\frac{1}{2J} \epsilon_{k\ell m}\epsilon_{jpq}
g_{i | k} f_{p | \ell} f_{q | m}\;\;.
\end{equation}
The gradient of the inverse transformation from Lagrangian to Eulerian coordinates,
${\bf h}={\bf f}^{-1}$, was expressed in terms of $\bf f$ through the algebraic relationship 
\begin{equation}
\label{inverseJacobian}
 \frac{\partial}{\partial x_j}h_i  = J_{ij}^{-1} = 
ad(J_{ij})J^{-1} = \frac{1}{2J}\epsilon_{ik\ell}
\epsilon_{j m n} f_{m | k}f_{n | \ell}\;\;.
\end{equation}
For the field equations we obtain with (\ref{g-gradient})
the following set of four Lagrangian  equations
(Buchert \& G\"otz 1987 for $\Lambda = 0$,
and Buchert 1989 for $\Lambda \ne 0$) ($i,j,k=1,2,3$ with cyclic
ordering; summation over repeated indices is understood):
\begin{eqnarray}
\label{lag1}
\frac{1}{2}\epsilon_{abc}\;\frac{\partial({g}_a,f_b,f_c)
}{\partial (X_1,X_2,X_3)} \; - \Lambda \,J
\; = \; - 4 \pi G \, \varrho_0 ({\bf X})\; ;\\
\label{lag2}
\epsilon_{pq \lbrack j} \frac{\partial ({g}_{i\rbrack},
f_p,f_q)}{\partial(X_1,X_2,X_3)} = 0 \;\;\;,\; i \ne j \;\;.\qquad\;\;\qquad
\end{eqnarray}
In the case of a {\it dust continuum}, 
the above equations form a set of four {\it evolution equations}, the 
{\it Lagrange--Newton system of equations}, 
by virtue of ${\bf g} = {\ddot{\bf f}}$.
Alternative forms of these equations may be found in Buchert (1996) and
Ehlers \& Buchert (1997).

The above equations not only hold for a continuum of {\it dust}; they simply represent a
transformation of the field equations, which also hold in the more general setting discussed 
in Sect. 2. The difference comes in when we represent the field strength $\bf g$ in terms
of the trajectory field; e.g. including isotropic pressure forces
the resulting Lagrangian evolution equations are investigated in Adler \& Buchert (1999).

Proposition~2 in Buchert (2006a) gives the following integral of the first 
Lagrangian equation (including its proof): 
\begin{equation}
\label{integralb}
{{\bf g}}^{I\Lambda}  =  ({\bf C} \cdot \nabla_{0}) {\bf f} \;J^{-1}\;
+\frac{\Lambda}{d}\,{\bf f} \;\;,\;\;{\bf C}:={\bf G} - \frac{\Lambda}{d}\,{\bf X}\;\;.
\end{equation}
Since this result is exact, we are entitled to apply the
transformation to comoving coordinates and peculiar--fields to this result.
Since we can adopt the same Lagrangian coordinates $\bf X$ in both cases in view of 
${\bf x}= a(t){\bf q}$, $a(t_0) = 1$ and since the background field strength evolves
as ${\bf g}_H = {\ddot a}{\bf q}$, 
we make the ansatz ${\bf g}= {\ddot a}{\bf F} + {\bf w}({\bf X},t)$,  
${\bf G}= {\ddot a}(t_0){\bf X} + {\bf W}({\bf X})$ and use Friedmann's equation 
$3\ddot a = \Lambda - 4\pi G a \varrho_H$ to obtain the integral ${\bf w}^{I}$ in
Eq.~(\ref{w_exactintegral})\footnote[6]{For an independent proof we could instead insert (\ref{w_exactintegral})
into the transformed Lagrangian equations given in Appendix A of Buchert (1989).}.

\section{The exact integral and its implications}
\label{sec:implications}

In order to learn more about the implications of the integral (\ref{w_exactintegral}),
we  now recover known model equations in cosmology (e.g. the 
Lagrangian perturbation scheme; this should be possible for the longitudinal parts, since
the integral (\ref{w_exactintegral}) is general in this case).
Then, we elaborate on possible applications.

\subsection{Recovering known cosmological models}
\label{subsec:implications_recovering}

We first write the general expression for $\bf w$ in terms of $\bf P$. In the case of a 
{\it dust continuum},
${\bf w} = a({\ddot{\bf P}} +2 H {\dot{\bf P}})$ (from Euler's equation for {\it dust}
${\bf w} = {\dot{\bf u}} + H{\bf u}$ with ${\bf u} = a {\dot{\bf P}}$),
we rewrite the integral (\ref{w_exactintegral}) in terms of $\bf P$ (we drop the 
index $I$ to ease the notation, consider the case $d=3$ and divide by $a$):
\begin{equation}
{\ddot {\bf P}} + 2 H {\dot {\bf P}} \,=\,
\frac{\left(\,{\bf K}\cdot \nabla_{0}\,\right)\,({\bf X}+{\bf P})}
{a^3 J_F}+\frac{4\pi G\varrho_H}{3}({\bf X}+ {\bf P})\;\;,
\end{equation}
with the following general expresssion for the Jacobian $J_F = a^{-3} J$:
\begin{equation}
\label{Jacobian}
J_F = 1 +  I (P_{i|k}) + II(P_{i|k})+III(P_{i|k})\;\;,
\end{equation}
where $I$, $II$ and $III$ denote the principal scalar invariants of the tensor in brackets.

Let us first expand the expression $1/J_F$ with the Jacobian (\ref{Jacobian}) to first order, 
$1/J_F = 1/(1 + \nabla_{0}
\cdot {\bf P} + ...)  \,\approx\,1 -   \nabla_{0}\cdot {\bf P}$.
We insert ${\bf K} =: {\bf W}- \frac{4\pi G \varrho_H (t_0)}{3}{\bf X}$,
drop terms that are quadratic in the fields ${\bf P}$ and ${\bf W}$, and 
replace the term 
$({\bf X}\cdot \nabla_{0}){\bf P} -
{\bf X} (\nabla_{0}\cdot {\bf P})$ by 
the (up to a transversal part equivalent) term 
$({\bf P}\cdot \nabla_{0}){\bf X} -
{\bf P} (\nabla_{0}\cdot {\bf X})\, =\, -2{\bf P}$.
This approximated equation then reads:
\begin{equation}
\label{firstorderdust}
{\ddot {\bf P}}^{(1)} + 2 H {\dot {\bf P}}^{(1)} - 4\pi G\varrho_H {\bf P}^{(1)}\,=\,
\frac{\bf W}{a^3}\;\;.
\end{equation}
Note that the factor $3$ disappeared from the term in front of ${\bf P}^{(1)}$.
This is the equation for longitudinal first--order Lagrangian perturbations (Buchert 1989, 1992).
For initial quasi--homogeneity, which is sometimes assumed, 
{\it cf.} Footnote 4, the right--hand side of this equation drops.

Expanding $1/J_F$ to higher orders soon yields messy expressions; e.g. by keeping only terms
quadratic in ${\bf P}^{(1)}$ and $\bf W$, we deal with an approximate equation of the 
form 
\begin{eqnarray}
{\ddot {\bf P}}^{(2)} + 2 H {\dot {\bf P}}^{(2)}- 4\pi G\varrho_H {\bf P}^{(2)} \,=\, 
\frac{\bf W}{a^3}\qquad\quad\nonumber\\
\qquad\qquad+\frac{1}{a^3}\left(
{\bf W}\cdot\nabla_{0}{\bf P}^{(1)} - {\bf W} \nabla_{0}\cdot {\bf P}^{(1)}\right)
\,+\,\cdots\;\;.
\end{eqnarray}
The last term in the above equation illustrates that we recover source terms for the
second--order perturbation solution ${\bf P}^{(2)}$, which correspond to the 
{\it local parts} (as introduced in Buchert 1993, Buchert 1994); the full terms including 
the non--local parts are solutions of
Poisson equations with sources given by the divergence of expressions of this type.

By expressing ${\bf w}$ through more general Euler equations that include presssure terms,
velocity dispersion, or deviations from mean field gravity ({\it cf.} Buchert \& Dom\'\i nguez 2005),
we can also expand the integral in order to recover known approximations, 
e.g., the first--order Lagrangian equation for longitudinal perturbations in a medium 
supported by isotropic pressure forces (Adler \& Buchert 1999).

\subsection{Iteration approach}
\label{subsec:implications_iteration}

A powerful possibility of applying the exact integral focusses on an iterative
definition of the trajectories or of the displacement field.
This iterative view exploits the fact that we know the field strength exactly (and in 
general with respect to the longitudinal field equation) along any
trajectory field that we could imagine.

We now exemplify the iteration procedure for a continuum of {\it dust}. Note that 
the integral (\ref{w_exactintegral}) can be viewed as a set of three partial differential 
equations for the trajectory field after inserting the general expression for the 
peculiar--acceleration field, e.g. for {\it dust}: 
${\bf w} = a({\ddot{\bf F}} + 2H {\dot{\bf F}})$.
Holding the Lagrangian coordinates fixed, i.e. for a single trajectory, we are dealing
with ordinary differential equations. 

The way we define the iteration scheme enjoys some freedom (non--unique definition).
We argue below for the following choice
of writing the integral (\ref{w_exactintegral}) as an iteration scheme: 
\begin{equation}
\label{w_exactiteration}
\frac{1}{a}{\bf w}^{[n+1]} - 4 \pi G\varrho_H {\bf F}^{[n+1]}  =  
\frac{\left(\,{\bf K} \cdot \nabla_{0}\,\right) {\bf F}^{[n]}}{a^3 J_F^{[n]} } - 
\frac{8 \pi G\varrho_H}{3}{\bf F}^{[n]} \;,
\end{equation}
with ${\bf K}$ given in (\ref{w_exactintegral}), and
$\frac{1}{a}{\bf w}^{[n+1]} 
= {\ddot{\bf F}}^{[n+1]} + 2H {\dot{\bf F}}^{[n+1]}$,  denoting the iteration steps 
in brackets to avoid confusion with the perturbation index used earlier.
In this way we have split the linear term $\frac{4 \pi G\varrho_H}{d}{\bf F}$ for 
$d\equiv 3$ into $4 \pi G\varrho_H - \frac{2}{3}4 \pi G\varrho_H$, the first term
we consider as belonging to $[n+1]$ and the second to $[n]$.
The reason for this choice becomes obvious after the  remarks below.

For the displacement field ${\bf P}={\bf F} - {\bf X}$, we obtain 
the following iteration scheme:
\begin{eqnarray}
\label{w_exactiterationP}
{\ddot {\bf P}}^{[n+1]} + 2 H {\dot {\bf P}}^{[n+1]} - 4\pi G\varrho_H {\bf P}^{[n+1]}\,=\,
\nonumber\\ \quad
\frac{4\pi G \varrho_H}{3}\left(\,{\bf X} - 2{\bf P}^{[n]}\,\right)\;+\;
\frac{\left(\,{\bf K} \cdot \nabla_{0}\,\right) ({\bf X}+{\bf P}^{[n]})}{a^3 
\det \left(\delta_{ik} + P_{i|k}^{[n]}\right)} \;\;.
\end{eqnarray}
Iteration consists in the strategy of feeding in a comoving displacement field
${\bf P}^{[n]}$ on the right--hand side of Eq.~(\ref{w_exactiterationP}) in order to obtain another 
displacement 
field ${\bf P}^{[n+1]}$ by solving second--order, ordinary differential equations 
for each fluid element $\bf X$.
The above choice of the iteration scheme is motivated by three arguments. First,
starting the iteration (\ref{w_exactiteration}) with the trivial trajectory field 
${\bf F}^{[0]}= {\bf X}$, which 
corresponds to a straight Hubble expansion, we obtain the equation governing 
the longitudinal first--order Lagrangian approximation exactly, ${\bf F}^{[1]}= {\bf X}+
{\bf P}^{[1]}$, with ${\bf P}^{[1]}$ obeying Eq. (\ref{firstorderdust}) for ${\bf P}^{(1)}$.
Second, the choice is supported by the fact that ${\bf F}^{[0]}$ (as a special exact solution of
the Lagrange--Newton system $\lbrace$\ref{lag1},\ref{lag2}$\rbrace$) 
also produces a special exact solution ${\bf F}^{[1]}$
of the same system (Buchert 1989). 
Third, to all orders in Lagrangian perturbation theory, 
the longitudinal differential operators for ${\bf P}^{(n+1)}$ 
have all the form of the differential operator for  ${\bf P}^{(1)}$, 
Eq.~(\ref{firstorderdust}), with sources involving the lower--order 
perturbation solutions (see Ehlers \& Buchert 1997, Sect.~3.2, 
for the general perturbation and solution schemes); compare with the left--hand side of 
Eq.~(\ref{w_exactiterationP}).

However, this does not mean that further iteration eventually produces further
exact solutions, and also that another choice could not perform as well as the 
above choice.
In order to illustrate the iteration scheme further, we compute the equation for the
second iterate ${\bf P}^{[2]}$. While the first iterate (insert ${\bf P}^{[0]} = {\bf 0}$ into
Eq.~(\ref{w_exactiterationP})) obeys
\begin{equation}
\label{firstiterate}
{\ddot {\bf P}}^{[1]} + 2 H {\dot {\bf P}}^{[1]} - 4\pi G\varrho_H {\bf P}^{[1]}\,=\,
\frac{\bf W}{a^3}\;\;,
\end{equation}
the second iterate may be found by inserting only a subclass of the general solution to 
Eq.~(\ref{firstiterate}) that corresponds to Zel'dovich's approximation (Zel'dovich 1970,
1973). After factorizing  ${\bf P}^{[1]}$ into time--dependent
functions and vector--functions of initial data, the time--dependent solutions consist of
two homogeneous  and one particular solution 
of the following equation  (Buchert 1992):
\begin{subequations}
\begin{equation}
\label{modes}
{\ddot\xi}(t) + 2 H(t) {\dot\xi}(t) - 4\pi G \varrho_H (t) (\xi (t)+1) \;=\;0\;\;,
\end{equation}
where $H(t)= \frac{\dot a}{a}$ and $\varrho_H (t) = \varrho_H (t_0)a^{-3}$ have to 
be expressed through solutions of Friedmann's differential equation:
\begin{equation}
\label{friedmann2}
\frac{{\dot a}^2}{a^2} - \frac{8\pi G \varrho_H}{3} + \frac{k}{a^2} - \frac{\Lambda}{3}
\;=\;0\;\;.
\end{equation}
\end{subequations}
Explicit forms of the functions $\xi (t)$ including a cosmological constant 
can be found in (Bildhauer et al. 1992; see also the supplement by 
Chernin et al. 2003).
We choose the restricted initial data set for which ${\bf U} \propto {\bf W}$ 
(${\bf U}({\bf X}):={\bf u}({\bf q},t_0)$) (for more details see: Buchert 1992). 
Inserting this subclass of first--order solutions (Zel'dovich's approximation),
${\bf P}^{[1Z]}:=b(t)\nabla_{0} \Psi ({\bf X})$, with the growing mode solution
$\xi_1 (t)=:b(t)$ of Eq.~(\ref{modes}), 
we can find the second iterate ${\bf P}^{[2Z]}$ corresponding to this restricted choice
as follows (we also use expression (\ref{Jacobian})):
\begin{eqnarray}
\label{seconditerate}
{\ddot {\bf P}}^{[2Z]} + 2 H {\dot {\bf P}}^{[2Z]} - 4\pi G\varrho_H {\bf P}^{[2Z]}\,=\,
\qquad\nonumber\\ 
\frac{4\pi G \varrho_H}{3}\left(\,{\bf X} - 2 b(t)\nabla_{0} \Psi ({\bf X})\,\right)\;+\;
\nonumber\\
\frac{\left(\,{\bf K} \cdot \nabla_{0}\,\right) ({\bf X}+b(t)\nabla_{0} 
\Psi ({\bf X}))}{a^3 
\left[\,1 + b I(\Psi_{|ik}) + b^2 II(\Psi_{|ik}) + b^3 III(\Psi_{|ik})\,\right]} \;\;.
\end{eqnarray}
This equation defines a set of ordinary differential equations parametrized by ${\bf X}$.
Its analytical solution (the subject of a forthcoming work) 
provides a test case for a numerical iteration scheme. For this
purpose we now give some useful technicalities.

\subsection{Numerical implementation}
\label{subsec:implications_numerics}

For the purpose of numerically implementing the iteration scheme (\ref{w_exactiterationP}),
we may rescale the dependent and independent variables as follows.
Following Shandarin (1980) except that we refer all quantities with subscript ``0'' to the initial
time $t_0$, we introduce the dimensionless and appropriately
scaled variables (Buchert 1989, App.A)
\begin{subequations}
\begin{eqnarray}
\label{scaledvariables}
{\bf\tilde q}:={\bf q}/q_0 \;\;;\;\;{\bf\tilde u}:=({\bf u}/ u_0 ) a\;\;;\;\; 
{\bf\tilde w}:=({\bf w}/w_0 ) a^3 \;\;,\nonumber\\
u_0 = q_0 / t_0\;\;;\;\;w_0 = q_0 / t_0^2 \;\;;
\;\;{\tilde\varrho}:=(\varrho / \varrho_H (t_0))a^3\;\;,
\end{eqnarray}
and a conformal transformation of the time--variable:
\begin{equation}
\label{scaledtime}
dT:= \frac{1}{t_0}\frac{dt}{a^2 (t)}\;\;,
\end{equation}
which is negative and tends to $-\infty$ at the Big--Bang singularity. With the help of 
(\ref{scaledtime}),
we can write the solutions of Friedmann's differential equation (\ref{friedmann2})
and the mass density parameter for the cases $\Lambda =0$\footnote[7]{For the cases
$\Lambda \ne 0$ we have to employ other strategies, e.g.: Bildhauer et al. 1992.}
in simple forms:
\begin{equation}
\label{scaledfriedmann}
a(T) = \frac{T_0^2 + k}{T^2 +k}\;\;\;;\;\;\;\Omega^m := \frac{8\pi G\varrho_H}{3H^2}=
\frac{T^2 +k}{T^2}\;\;,
\end{equation}
with $k = 0, \pm 1$. For an Einstein--de Sitter model ($k=0$), we have $T_0 = -3$, and for 
the other models $T_0 = - \sqrt{\frac{k}{\Omega^m_0 -1}}$.
Below we also use the relations $4\pi G\varrho_H (t_0) = \frac{3}{2}\Omega^m_0 
H_0^2$ and $4\pi G\varrho_H (t_0)t_0^2 = 6/ (T_0^2 +k)$;\\
(for further details see: Buchert 1989, App.A).

For the scaled displacement field, 
\begin{equation}
{\bf\tilde P}:= \frac{{\bf P}}{q_0}\;\;;\;\;{\bf\tilde X}:= \frac{{\bf X}}{q_0}\;\;;\;\;
{\tilde\nabla}_{\bf 0} = q_0 \,\nabla_{0} \;\;,
\end{equation}
\end{subequations}
Eq.~(\ref{w_exactiterationP})
can be rewritten as a set of first--order (for each ${\bf\tilde X}$ ordinary)
differential equations:
\begin{eqnarray}
\label{w_exactiterationPrescaled}
\frac{d}{dT}{\bf\tilde P}^{[n+1]}\;=\;{\bf\tilde u}^{[n+1]}\;\;; \qquad\qquad\qquad\qquad\nonumber\\
\frac{T^2 +k}{2}\,\frac{d}{dT} {\bf\tilde u}^{[n+1]}  - 3 \;{\bf\tilde P}^{[n+1]}\,=\;\qquad\qquad\qquad\quad
\nonumber\\ \quad
\left(\,{\bf\tilde X} - 2{\bf\tilde P}^{[n]}\,\right)\;+\;
\frac{\left(\,{\bf\tilde K}
\cdot {\tilde\nabla}_{\bf 0}\,\right) \left({\bf\tilde X}+{\bf\tilde P}^{[n]}\right)}{ 
\det \left(\delta_{ik} + {\tilde P}_{i|k}^{[n]}\right)} \;\;,
\end{eqnarray}
with ${\bf\tilde K}:=\frac{1}{2}(T_0^2 +k){\bf\tilde W} -{\bf\tilde X}$.
The differential operator in the second equation simplifies further by 
integrating along the time $\tau$ with $d\tau :=2 a(T) /(T_0^2 +k)\,dT$.
Initial data for $\bf\tilde W$ follow from solving the initial Poisson equation for 
the density contrast $\delta = {\tilde\varrho}-1$:
\begin{equation}
{\tilde\nabla}_{\bf 0} \cdot \frac{1}{2}(T_0^2 +k){\bf\tilde W} = -3\;\delta ({\bf\tilde X})\;\;;\;\;
{\tilde\nabla}_{\bf 0} \times {\bf\tilde W} = {\bf 0}\;\;.
\end{equation}
Initial data for $\bf\tilde U$ could be specified by the special choice 
${\bf\tilde U}={\bf\tilde W}$.

As a first test of a numerical scheme one chooses 
${\bf\tilde P}^{[0]} = {\bf 0}$, so that ${\bf\tilde P}^{[1]}$
should be identical to Zel'dovich's approximation (Zel'dovich 1970, 1973).
A second test will be provided by the analy\-tical solution for the second iterate, 
as mentioned above. A third test should measure the (artificial) vorticity of $\bf w$ as
introduced by the integral for generic initial data; here, one can compare with the exact 
expression (\ref{transverseintegral}) below for the transverse part of the field strength.

All peculiar--fields appearing in the calculations must be periodic on the
largest scale to assure
that the average model is Friedmannian. This construction is necessary for the uniqueness
of a Newtonian solution and implies a globally vanishing `backreaction' (see: Buchert \&
Ehlers 1997).

\section{Summary and prospects}
\label{sec:summary}

Current analytical models (excluding higher--order perturbative corrections)
basically follow from the simple assumption that the 
peculiar--gravitational field strength is proportional to the displacement field, 
Equation~(\ref{w_approx}):
${\bf w} = \frac{\bf W}{a^2}+4\pi G \varrho_H a{\bf P}$. Eq.~(\ref{P_approxEJN})
furnishes the up to date most general model equation based on this assumption.
More frequently employed models, like the celebrated {\it Zel'dovich approximation}
and the {\it adhesion approximation}, even imply the tighter restriction of proportionality
for the peculiar--gravitational field strength to the peculiar--velocity, which is 
a very good assumption in the weakly non--linear regime, but certainly fails in a highly
non--linear situation. Consequences of the 
above remarks have been investigated in detail in
(Buchert \& Dom\'\i nguez 2005).

We have argued that relationship (\ref{w_approx}) needs generalization for the
understanding of the non--perturbative regime of cosmic structure formation.
Integrating the transport equation for the peculiar--gravitational field strength,
Eq.~(\ref{w_transport}), we obtained an exact integral, generalizing 
relationship (\ref{w_approx}). It was demonstrated that this integral can be employed
to define an iteration scheme that allows us to obtain the peculiar--gravitational
field strength for any given family of trajectories, which  solves the
Lagrangian evolution equation corresponding to ${\nabla}_{\bf q}\cdot{\bf w} =
- 4\pi G \varrho_H a \delta$ {\it in general}.
This property, together with the experience of the good performance of 
Lagrangian perturbation schemes based on the longitudinal part only 
(e.g., Buchert et al. 1997), supports the expectation that the integral and its
corresponding iteration scheme provide a powerful approximation for the non--perturbative
regime of structure formation. 

The qualitative difference from a Lagrangian perturbation analysis is due to the fact that
the former lacks the important leading term proportional to the density in 
(\ref{w_exactintegral}). Lagrangian perturbation solutions predict a smooth gravitational
field strength when crossing caustics in the density field for all orders in perturbation theory
(compare the general perturbation and solution schemes given by Ehlers \& Buchert 1997). 
Therefore, we are entitled to consider the integral (\ref{w_exactintegral}) as a genuinly
non--perturbative result. It shows that 
a blow--up of the field strength at caustics is a generic property of the gravitational
collapse. Counterarguments 
based on exact solutions with symmetry do not apply to the generic situation. 
Consider, as an example,
plane--symmetric motions on a three--dimensional homogeneous--isotropic
background. In that case the general exact solution (at the same time a solution
of Eq.~(\ref{firstorderdust})) does
not predict a singular field strength: specifying the integral (\ref{w_exactintegral}) to 
plane--symmetric motion, we infer from the expression proportional to the density,
\begin{equation}
\label{planesymmetry}
\frac{\left(\,({\bf W}-\frac{4\pi G\varrho_H (t_0)}{3}{\bf X}) \cdot 
\nabla_{0}\,\right) ({\bf X}+{\bf P})}{a^2 J_F}\;\;, 
\end{equation}
\begin{equation}
\qquad{\rm with}\;\;J_F = 1 + P_{1|1}\;,
\end{equation}
that the Jacobian cancels the directional derivative term exactly, and 
$w_1^{\rm \; plane}=4\pi G\varrho_H a P_1 + W_1 /a^2$ 
does not blow up at the caustic where $ 1 + P_{1|1}=0$.

One of the most promising application fields of the integral and its iteration scheme
could be the following.
Since (\ref{w_exactintegral}) provides $\bf w$ as a local functional of  the
displacement field $\bf P$, it may substantially enhance the power of reconstruction 
methods (Croft \& Gazta{\~n}aga 1997; Susperregi \& Buchert 1997; 
Courteau \& Willick 2000, Courteau \& Dekel 2001;
Brenier et al. 2003, Mohayaee et al. 2003). 
As mentioned above, the integral (\ref{w_exactintegral}) generalizes the assumption
(\ref{w_approx}), which itself lies at the basis of most analytical models 
(as summarized in this paper) and which already
furnishes a more general ansatz for reconstruction methods, compared to those
that are currently implemented, e.g. those that assume the proportionality of $\bf w$
and $\bf u$.

To propose a non--perturbative approximation based on the integral 
(\ref{w_exactintegral}) is  also 
motivated by simplicity,  besides the advantage of a local expression.
On theoretical grounds, however, there are limitations to the 
integral  (\ref{w_exactintegral}), because it will produce artificial vorticity of $\bf w$,
since a generic approximation based on (\ref{w_exactintegral}) will not satisfy 
${\nabla}_{\bf q}\times {\bf w} = {\bf 0}$. Numerical work has to confirm the
expectation that the longitudinal part of the field equations provides the dominant
contribution. The transverse contribution and the contribution from non--local terms
can also be quantified, based on results of a forthcoming work on transported 
differential forms associated with the gravitational field strength; we here already give 
the exact Lagrangian integral that solves the transverse
field equation {\it in general} (Buchert 2006b):
\begin{equation}
\label{transverseintegral}
{\bf w}^{\rm transverse}_k  =  \frac{1}{a J_F} 
\left( W_i J_{ik}^{\rm sub} - ( {\bf W} \times 
\lbrack ({\nabla}_0 \times {\bf F} ) \cdot {\nabla}_0 \;{\bf F} 
\rbrack)_k \right),
\end{equation}
where $J_{ik}^{\rm sub}$ denotes the subdeterminants of $(F_{i | k})$.

Since a general integral, if it exists, would include
non--local parts, and thus would require the solution of elliptic boundary value problems
at all times, it  may be of limited practical use.

\begin{acknowledgements}

  This work was supported by the ``Sonderforschungsbereich SFB 375 f\"ur
  Astro--Teilchenphysik der Deutschen Forschungsgemeinschaft''.
  Thanks go to St\'ephane Colombi for stimulating discussions and
  thoughts on the numerical implementation of the iteration scheme.
\end{acknowledgements}

\vspace{-20pt}

\section*{References}

\def\refitem{\par\noindent\hangindent\parindent\hangafter1}

\refitem
Adler, S., \& Buchert, T. 1999, A\&A, 343, 317
\refitem
Bagla, J.S., \& Padmanabhan, T. 1997,  Pramana, 49, 161--192 (astro--ph/0411730)
\refitem
Bernardeau, F., Colombi, S., Gazta\~naga, E., \& Scoccimarro, R. 2002, 
Phys. Rep., 367, 1
\refitem
Bertschinger, E. 1998, ARA\&A, 36, 599
\refitem
Bildhauer, S., Buchert, T., \& Kasai, M.\ 1992, A\&A, 263, 23 
\refitem
Binney, J., \& Tremaine, S. 1987, Galactic Dynamics, Princeton
University Press
\refitem
Bouchet, F.R., Juszkiewicz, R., Colombi, S. and Pellat, R. 1992, ApJ, 394, L5
\refitem
Bouchet, F.R., Colombi, S., Hivon, E. and Juszkiewicz, R. 1995, A\&A, 296, 575 
\refitem
Brenier, Y., Frisch, U., H\'enon, M., Loeper, G., Matarrese, 
S., Mohayaee, R., \& Sobolevskii, A.\ 2003, MNRAS, 346, 501
\refitem
Buchert, T. 1989, A\&A, 223, 9
\refitem
Buchert, T. 1992, MNRAS, 254, 729
\refitem
Buchert, T. 1993, A\&A, 267, L51
\refitem
Buchert, T. 1994, MNRAS, 267, 811
\refitem
Buchert, T. 1996 In Proc. IOP `Enrico Fermi', Course CXXXII 
(Dark Matter in the Universe), Varenna 1995, eds.
S. Bonometto, J. Primack, \& A. Provenzale, IOS Press Amsterdam, pp. 543-564
\refitem
Buchert, T. 2006a, Phys. Lett. A, 354, 8
\refitem
Buchert, T. 2006b, work in preparation
\refitem
Buchert, T., \& Dom\'\i nguez, A. 1998, A\&A, 335, 395
\refitem
Buchert, T., \& Dom\'\i nguez, A. 2005, A\&A, 438, 443
\refitem
Buchert, T., \& Ehlers, J. 1993, MNRAS, 264, 375
\refitem
Buchert, T., \& Ehlers, J. 1997, A\&A, 320, 1
\refitem
Buchert, T., \& G\"otz, G. 1987, J. Math. Phys., 28, 2714
\refitem
Buchert, T., Karakatsanis, G., Klaffl, R., \& Schiller, P.\ 1997, 
A\&A, 318, 1
\refitem
Catelan, P. 1995, MNRAS, 276, 115
\refitem
Chernin,  A.D., Nagirner, D.I., \& Starikova, S.V.\ 2003, A\&A, 399, 19
\refitem
Courteau, S. \& Dekel, A. 2001, in: {\it Astrophysical Ages and Times Scales}, 
ASP Conference Series Vol. 245., T. v. Hippel, C. Simpson, N. Manset (eds.)
\refitem
Courteau, S. \& Willick, J. 2000 (eds.), {\it Cosmic Flows Workshop}, 
ASP Conference Series, Vol. 201.
\refitem
Croft, R.A.C., \& Gazta{\~n}aga, E.\ 1997, MNRAS, 285, 793
\refitem
Ehlers, J., \& Buchert, T. 1997, Gen. Rel. Grav., 29, 733
\refitem
Gabrielli, A., Baertschiger, T., Joyce, M., Marcos, B. \& Sylos--Labini, F. 2006; cond--mat/0603124
\refitem
Gurbatov, S.N., Saichev, A.I., \& Shandarin, S.F. 1989, MNRAS, 236, 385
\refitem
Ma, C.--P., \& Bertschinger, E. 2004, Ap.J. 612, 28
\refitem
Morita, M., \& Tatekawa, T. 2001, MNRAS, 328, 815
\refitem
Mohayaee, R., Frisch, U., Matarrese, S., \& Sobolevskii, A 2003, 
A\&A, 406, 393
\refitem
Moutarde, F., Alimi, J.--M., Bouchet, F.R., Pellat, R., \& Ramani, A. 1991,
Ap.J. 382, 377
\refitem
Peebles, P.J.E. 1980, {\it The Large--scale Structure of the Universe},
Princeton Univ.~Press 
\refitem
Sahni, V., \& Coles, P. 1995, Phys. Rep. , 262, 1
\refitem
Shandarin S.F. 1980, Astrofizika 16, 769; 1981: Astrophysics 16, 439
\refitem
Serrin, I. 1959 In Encyclopedia of Physics, Vol. III.1, ed. S. Fl\"ugge,
Springer Berlin
\refitem
Shirokov, A., \& Bertschinger, E. 2006, submitted to Ap.J.Suppl.; astro--ph/0505087
\refitem
Steinmetz, M., 2003, in: {\it Hubble's Science Legacy: Future Optical/Ultraviolet 
Astronomy from Space}, ASP Conference Proceedings, K.R. Sembach et al. (eds.) 
Vol. 291, p.237
\refitem
Susperregi, M., \& Buchert, T. 1997, A \& A 323, 295 
\refitem
Tatekawa, T., Suda, M., Maeda, K., Morita, M., \& Anzai, H. 2002, Phys. Rev. D66, 064014
\refitem
Tatekawa, T. 2004a, Phys. Rev. D69, 084020
\refitem
Tatekawa, T. 2004b, Phys. Rev. D70, 064010
\refitem
Tatekawa, T. 2005a, Phys. Rev. D71, 044024
\refitem
Tatekawa, T. 2005b, Phys. Rev. D72, 024005
\refitem
Vanselow, M. 1995, Diploma Thesis, Ludwig--Maximi\-lians--Universit\"at M\"unchen
(in German).
\refitem
Zel'dovich, Ya.B. 1970, A\&A, 5, 84
\refitem
Zel'dovich, Ya.B. 1973, Astrophysics, 6, 164

\end{document}